 \documentclass[useAMS,usenatbib,usegraphicx]{mn2e}
\usepackage{epsfig}
\usepackage{amsmath} %need for line break in equ 3
\usepackage{rotating}           % for sideways tables/figures
\usepackage{color}     
\usepackage{graphicx}
\usepackage{times}
\usepackage{upgreek} % for up $\uptau$ 

\def\kms{km ${\rm s}^{-1}$}

\def\ch2{$\chi^2$}
\def\dg{$^{\circ}$}
\def\kms {\hbox{${\rm km\ s}^{-1}$}}

\def \AL {$\alpha $}     %  gr. alpha
\def \HI {H{\sc \,i}}
\def \WpHz {W Hz$^{-1}$}
\def\lapp{\ifmmode\stackrel{<}{_{\sim}}\else$\stackrel{<}{_{\sim}}$\fi}
\def\gapp{\ifmmode\stackrel{>}{_{\sim}}\else$\stackrel{>}{_{\sim}}$\fi}

\title[\HI\ column densities in compact radio sources]{On the \HI\  column density--radio source size anti-correlation in compact radio sources} \author[S. J. Curran et
al.]{S. J. Curran$^{1,2}$\thanks{E-mail:
    sjc@physics.usyd.edu.au}, J. R. Allison$^{1}$, M. Glowacki$^{1}$, M. T. Whiting$^{3}$ and  E. M. Sadler$^{1,2}$\\
  $^{1}$Sydney Institute for Astronomy, School of Physics, The University of Sydney, NSW 2006, Australia\\
  $^{2}$ARC Centre of Excellence for All-sky Astrophysics (CAASTRO)\\
  $^{3}$CSIRO Astronomy and Space Science, PO Box 76, Epping NSW 1710,
  Australia}
 
\begin{document}

\date{Accepted ---. Received ---; in original form ---}

\pagerange{\pageref{firstpage}--\pageref{lastpage}} \pubyear{2013}

\maketitle

\label{firstpage}

\begin{abstract}
  Existing studies of the atomic hydrogen gas content in distant galaxies, through the absorption of the 21-cm line,
  often infer that the total column density, $N_{\rm HI}$, is anti-correlated with the linear extent of the background
  radio source, $d_{\rm em}$. We investigate this interpretation, by dissecting the various parameters from which
  $N_{\rm HI}$ is derived, and find that the relationship is driven primarily by the observed optical depth, $\tau_{\rm
    obs}$, which, for a given absorber size, is anti-correlated with $d_{\rm em}$.  Therefore, the inferred $N_{\rm
    HI}-d_{\rm em}$ anti-correlation is merely the consequence of geometry, in conjunction with the assumption of a
  common spin temperature/covering factor ratio for each member of the sample, an assumption for which there is scant
  observational justification.  While geometry can explain the observed correlation, many radio sources comprise two
  radio lobes and so we model the projected area of a two component emitter intercepted by a foreground absorber. From
  this, the observed $\tau_{\rm obs} - d_{\rm em}$ relationship is best reproduced through models which approximate
  either of the two Fanaroff \& Riley classifications, although the observed scatter in the sample cannot be duplicated
  using a single deprojected radio source size. Furthermore, the trend is best reproduced using an absorber of diameter
  $\sim100 - 1000$~pc, which is also the range of values of $d_{\rm em}$ at which the 21-cm detection rate peaks. This
  may indicate that this is the characteristic linear size of the absorbing gas structure.
\end{abstract}
\begin{keywords}
  galaxies: fundamental parameters -- galaxies: active -- galaxies:
  evolution -- galaxies: ISM -- radio lines: galaxies
\end{keywords}

\section{Introduction}
\label{intro}

The 21-cm transition of hydrogen (\HI) probes the cool, star-forming, component of the neutral gas throughout the
Universe. Due to the low probability of the transition, compounded by the inverse square law, this is
difficult to detect in emission at redshifts of $z\gapp0.1$ with current instruments, although in absorption the line strength depends only upon the total hydrogen
column density and the strength of the background continuum (see Eq. \ref{enew}). At $z\gapp0.1$, this transition
has been detected along eighty different sight-lines, half of which occur in galaxies that intervene the sight-lines to more distant quasars
(see \citealt{gsp+12} and references therein), with the other half occuring within the host galaxy of the radio source
itself (associated absorption; compiled in \citealt{cw10}, with more recent results published in
\citealt{cwm+10,cwwa11,cwsb12,cwt+12}).

Of the associated absorbers, detection rates appear to be higher in those objects classified as
``compact''.\footnote{That is, gigahertz peaked spectrum sources (GPS, $\lapp1$ kpc) and compact steep spectrum sources (CSS, $\sim1- 10$ kpc),
in addition to their various sub-classes --  compact
  symmetric objects (CSO), compact flat spectrum (CFS), high frequency peaker galaxies (HFPs) and low power compact
  (LPC) radio sources (e.g. \citealt{ode98,fan00,omd06}).}  Furthermore, \citet{pcv03} reported an
anti-correlation between the derived atomic hydrogen column density, $N_{\rm HI}$, and the projected linear size of the
background radio emission, $d_{\rm em}$, a finding which has since been sustained 
through the addition of further data (e.g.  \citealt{gs06,gs06a,gss+06,omd06,css11}). 

\begin{figure}
\centering \includegraphics[angle=270,scale=0.70]{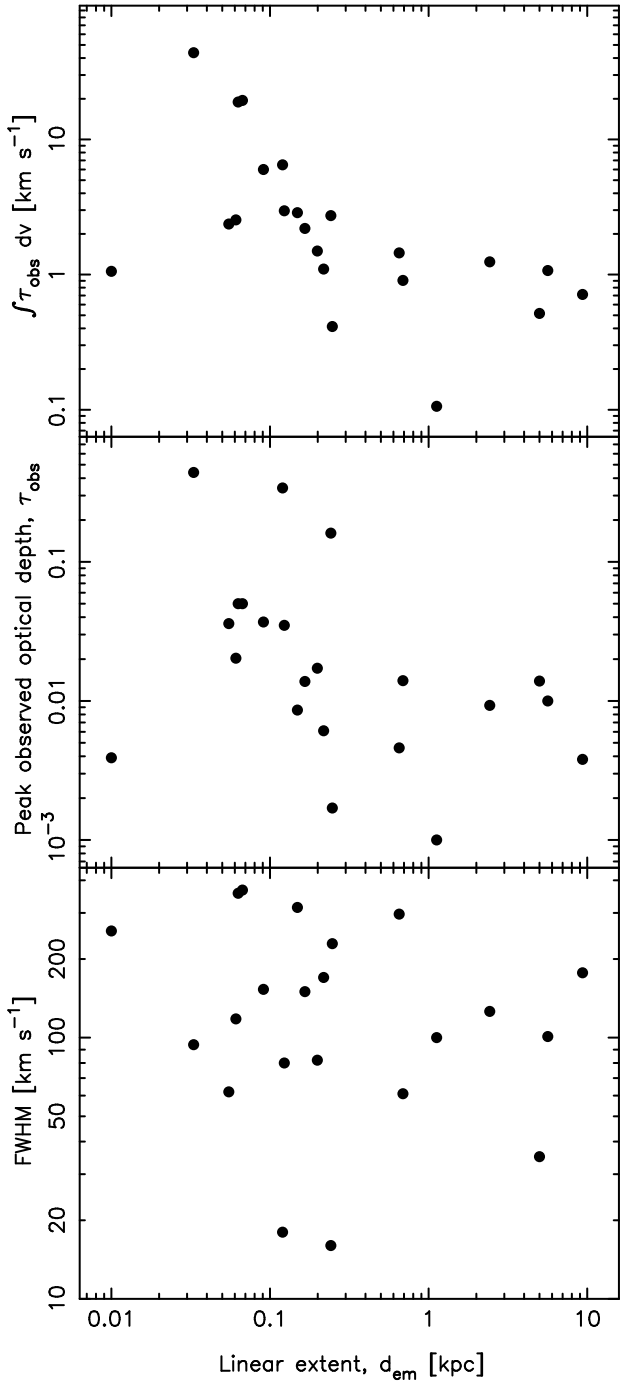}
\caption{The velocity integrated optical depth 
  (top), the peak observed optical depth (middle) and the full-width half maxima
  (bottom) of the \HI\ 21-cm line versus the projected linear size. From \citet{pcv03}.}
\label{pcv}
\end{figure}

This interpretation suggests that
the smaller the radio source, the denser the absorbing medium, which could be consistent with either the ``frustration scenario'' \citep{vmh89},
where the jets are confined, or the ``youth scenario''   (\citealt{ffd+95}), where the gas has yet to be expelled (supported by \citealt{pcv03}).
In either case,  there is one major flaw in the interpretation that the column density is anti-correlated with the jet size: It is 
the observed velocity integrated optical depth of the 21-cm line that is actually measured, rather than the \HI\ column density
(Fig.~\ref{pcv}, top), which is derived by assuming a spin temperature and covering factor (usually $T_{\rm spin} =
100$~K and $f=1$).
Where column densities are known\footnote{Usually in $z_{\rm abs}\gapp1.7$ damped Lyman-\AL\ absorption systems
  (DLAs).}, the observable quantity $T_{\rm spin}/f$ is seen to vary by a factor of at least $\sim170$ (from $ 60$~K,
\citealt{ctp+07} to 9950~K, \citealt{kc02}), and so this assumption cannot be justified.  Furthermore, due to the
scatter in the full-width half maxima of the 21-cm profiles (Fig. \ref{pcv}, bottom), \citeauthor{pcv03} acknowledge
that the correlation is driven by the observed optical depth (Fig.~\ref{pcv}, middle).  This therefore suggests that the
observed relationship arises directly from an {\em optical depth}--linear size anti-correlation, the possible cause of
which we explore in this paper. %letter.

\section{Optical depth and radio source size}

\subsection{Observed optical depth and covering factor}
\label{odcf}

The atomic hydrogen column density along the line-of-sight, $N_{\rm HI}$, is related to the velocity integrated
optical depth of the 21-cm absorption, $\int\!\tau\,dv$,
via \citep{wb75}:
\begin{equation}
N_{\rm HI}=1.823\times10^{18}\,T_{\rm spin}\int\!\tau\,dv\,,
\label{enew}
\end{equation}
where $T_{\rm spin}$ [K] is the mean harmonic spin temperature of the gas and the
optical depth is given by
\begin{equation}
\tau \equiv-\ln\left(1-\frac{\tau_{\rm obs}}{f}\right).
\label{tau_obs}
\end{equation}
Here $\tau_{\rm obs} \equiv\Delta S/S$ is the observed optical depth
of the line, where $\Delta S$ is the spectral line depth and $S$ the
continuum flux of the background source.  The covering factor, $f$,
quantifies how effectively the absorber intercepts the flux (ranging
from zero to unity)\footnote{In fact, from $\tau_{\rm obs}$ to unity for a uniform source, based upon the limit 
$f > \tau_{\rm obs}$ imposed by Eq. \ref{tau_obs} \citep{obg94}.}, meaning that $\tau_{\rm obs} \leq
\tau$. In the optically thin
regime, 
Eq.~\ref{tau_obs} simplifies to
 $\tau \approx\tau_{\rm obs}/{f}$, so that Eq.~\ref{enew} is approximated by
 $N_{\rm HI} \approx 1.823\times10^{18}\,(T_{\rm  spin}/f)\int\!\tau_{\rm obs}\,dv$.
 Therefore, with knowledge of the value of $T_{\rm spin}/f$, the column density can be obtained from the observed
 integrated optical depth. However, this ratio is generally unknown\footnote{Conversely, it is common practice to use
   $N_{\rm HI}$, where known from observations of the Lyman-\AL\ line, in conjunction with $\int\!\tau_{\rm obs}\,dv$,
   from the 21-cm line, to obtain $T_{\rm spin}/f$ in DLAs (See \citealt{cur12} and references therein).}, with the assumption of a single value being applied
to yield the column density in each source \citep{pcv03,gs06,gs06a,gss+06,css11}.

%Thus, it is immediately obvious that it is in fact 
We now consider, without evidence to the contrary, that the column density of a foreground
absorber is independent of the fraction of background radio flux it covers.
In the optically thin regime, for a given absorber
the observed optical depth is proportional to the covering factor (Eq. \ref{tau_obs}), which in turn
is anti-correlated with the size of the background source. That is,
%\begin{equation}
%\tau_{\rm obs}\propto f =\left(\frac{A_{\rm abs}}{A_{\rm em}}\right) = \left(\frac{d_{\rm abs}}{d_{\rm em}}\right)^2,
%\label{tau_prop}
%\end{equation}
$\tau_{\rm obs}\propto f =\left(A_{\rm abs}/A_{\rm em}\right),$
where the last term is the ratio of the cross-sectional area of the absorber to that of the background emitter (where
$z_{\rm abs} \approx z_{\rm em}$, cf. \citealt{cur12}).% and the last term the ratio of the respective linear sizes.

\begin{figure}
\centering \includegraphics[angle=270,scale=0.73]{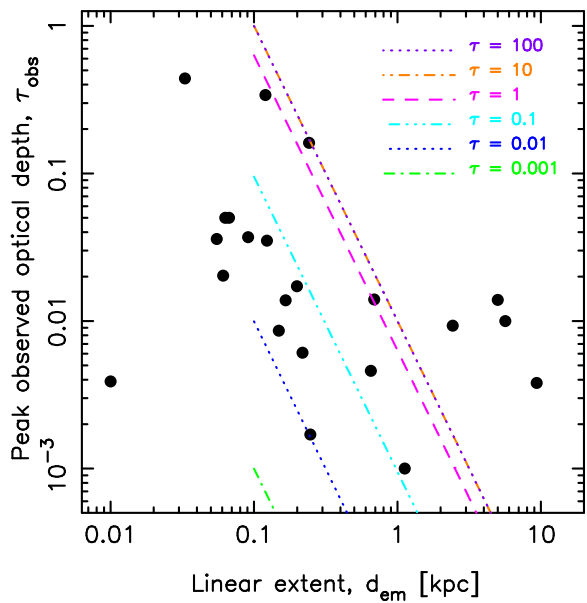} % mv pcv-tau-curves.ps pcv-tau-curves_squ.ps
\caption{The peak observed optical depth for the data of \citet{pcv03} [cf. Fig. \ref{pcv}, middle], but with the
overlain lines showing how, for a given absorber size, the optical
depth varies for $f = (d_{\rm abs}/d_{\rm em})^2$. 
Here the absorption cross-section is set to $d_{\rm abs} = 100$ pc.
The values of $\tau$, ranging from 0.001 to 100, are given by the (approximate, for $\tau\geq1$) maximum values on the ordinate.}
\label{pcv2}
\end{figure}

Therefore, based upon its definition alone, the observed optical depth is anti-correlated with the size of the radio
source. This is seen in a plot of these two quantities (Fig. \ref{pcv}, middle), with the scatter being due to
differences in the actual optical depths, $\tau$, and, possibly, the absorber sizes. %, $d_{\rm abs}$, which also feature in (Eq. \ref{tau_prop}).  
In Fig.~\ref{pcv2} we re-plot this and overlay the variation of $\tau_{\rm obs}$
with $d_{\rm em}$ for $\tau_{\rm obs} = f(1-e^{-\tau})$ and $f = (A_{\rm abs}/A_{\rm em}) = (d_{\rm abs}/d_{\rm em})^2$. In the optically thin approximation
(where $\tau_{\rm obs} \approx f\,\tau$), setting the absorber size arbitrarily small can spread the overlaid lines over all of the data points
(with equal spacing on the abscissa, as seen for $\tau_{\rm obs}\leq1$). This, however, is not justified for $\tau\gapp0.3$ and
using $\tau_{\rm obs} = f(1-e^{-\tau})$ introduces the compression at large optical depths since $\tau_{\rm obs}\rightarrow f$ as
$\tau \rightarrow \infty$.\footnote{Although $\tau_{\rm obs} \lapp0.3$ for the vast majority of absorbers.}  Increasing the size of the absorber (e.g. to $d_{\rm abs} \gapp 500$ pc) 
accounts for the four points with
$d_{\rm em} \gapp2$ kpc, but leaves the smaller sources unaccounted for, although, as stated above, there is no reason
to expect a common absorber size.

\begin{figure}
\centering \includegraphics[angle=270,scale=0.73]{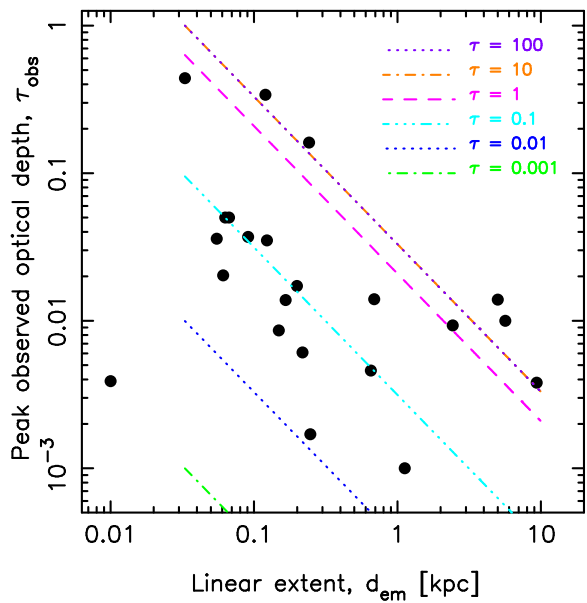} 
\caption{As Fig. \ref{pcv2}, but for $f = d_{\rm abs}/d_{\rm em}$ and $d_{\rm abs} = 33$ pc (the smallest of the bulk values of $d_{\rm em}$, i.e. excluding the outlier
at $d_{\rm em}\approx10$ pc). Again, the $\tau = 10$ and $100$ lines are coincident.}
\label{pcv1}
\end{figure}

The data are better fit by a covering factor which depends linearly on the source size, i.e. $f = d_{\rm abs}/d_{\rm em}$,
which can trace most of the points for a single absorber size  (Fig. \ref{pcv1}).
This suggests that the geometry of the emission could be dominant along a single axis. This may be expected since
high resolution imaging reveals the sources of emission to usually be two main lobes (e.g. \citealt{tkm+02,pcv03} and references therein),
with the extent of the radio emission being determined by the lobe separation. % (i.e. along a single axis). 
%However, a gap in the emission is introduced by the lobe spacing which  will also have an effect on the area of emitter intercepted by the absorber. 
We now explore this, in conjunction with the possibility
that the large range of emitter sizes may be due to projection effects, by extending the sample to also include the non-compact sources.

\subsection{Extending to the general population of radio sources}
\label{ext}

\begin{figure}
\centering \includegraphics[angle=270,scale=0.73]{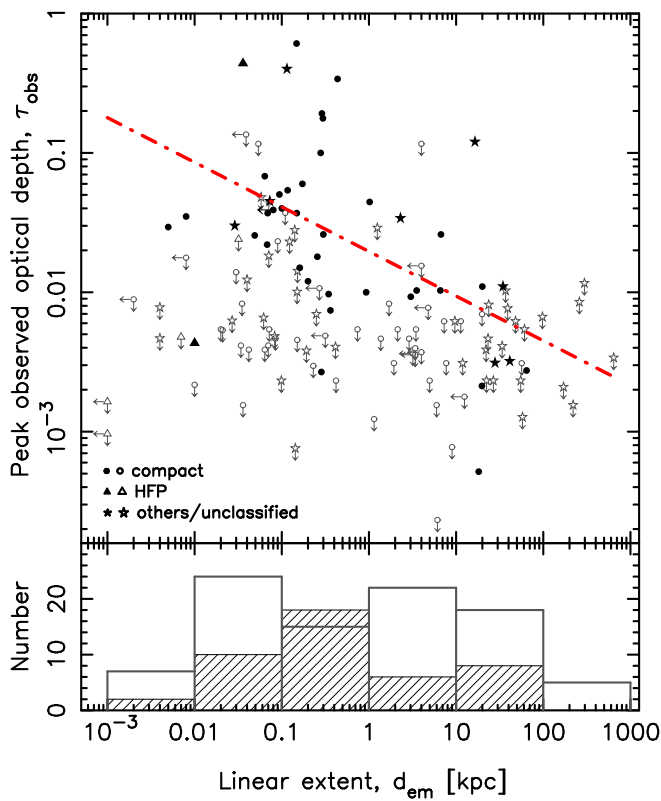}
\caption{The peak observed optical depth versus the projected linear size (where available) 
for all of the redshifted radio sources searched in 21-cm absorption to date (compiled in \citealt{cw10,ace+12}).The filled markers/hatched histogram represent the
detections and the unfilled markers/histogram show the $3\sigma$ upper limits to the optical depth for the non-detections at a 
spectral resolution of 167 \kms\ (see \citealt{cwsb12}). The circles designate compact objects and the stars 
non-compact/unclassified, with the triangles showing the high frequency peaker galaxies, which, although compact, \citet{omd06} argue do not follow the trend. 
The line shows the least-squares-fit to the detections. }
\label{all}
\end{figure}

In discussing the effects of radio source size on the strength of the 21-cm absorption, \cite{cw10} noted that the addition of
non-compact/unclassified sources did not dilute the anti-correlation. Furthermore, from optical, near and far-infrared observations,
no difference is found between compact and extended radio sources (\citealt{dob+98} and \citealt{fpf+00}, respectively).
Including the non-compact/unclassified objects for which $d_{\rm em}$ is available (e.g. \citealt{gss+06}), the
Kendall's $\mathcal{T}$ two-sided probability of the correlation between observed optical depth and projected source
size arising by chance is $P(\mathcal{T})=0.0044$. %\footnote{We have substituted tu ($\tau$) with $\mathcal{T}$ in order  to avoid confusion with optical depth.}  
This is in comparison to the $P(\mathcal{T})<0.01$ ``column density''
correlation \citep{pcv03} and corresponds to a significance of $2.84\sigma$, assuming Gaussian statistics
(Fig. \ref{all}).\footnote{This significance falls to $1.96\sigma$ when the non-detections are included (via the survival analysis of \citealt{lif92a}), since these are subject to
  additional effects \citep{cw10}.}
% 14 det_large.txt       for size_cut = 0, i.e. 1kpc
% 30 det_small.txt             => rate at < 1kpc (small) is 30/30+47 = 0.39
% 47 non_small.txt                              > 1kpc (large) is 14/14+46 = 0.23
% 46 tau_size.txt

\citet{gss+06} state that the incidence of \HI\ absorption is much higher for compact sources than for extended
sources. The histogram in Fig. \ref{all} shows the relative number of detections (hatched) and non-detections (unfilled)
in each $\log_{10}(d_{\rm em})$ bin. One thing immediately obvious from this is that, while the detection rates are
$\lapp30$\% in all other bins, between 0.1 and 1 kpc the rate exceeds 50\%. A peak in the observed optical
depth is also apparent in this bin (the top panel of Fig. \ref{all}). That is, there may exist a ``resonance'' between the absorber and radio source size. %($d_{\rm abs}$ \& $d_{\rm em}$, respectively). 
This could arise if when $d_{\rm abs} \ll d_{\rm em}$ the covering factor is low, whereas when
$d_{\rm abs} \gg d_{\rm em}$,  %if the absorber is clumpy,
 the chance of a misalignment between the absorber and emitter along our sight-line
is greater for a smaller emission region, indicative of a clumpy absorbing medium.
%This would therefore suggest that the 21-cm absorbing cross-sections rarely deviate far from $d_{\rm abs} \sim100-1000$ pc over a wide range of environments and morphologies.  
Interestingly, $\sim100$~pc is the scale of the densest \HI\ clouds found from low redshift emission studies \citep{bra12}.

We find the 21-cm detection rate to be 39\% (30/77) at $d_{\rm em}<1$ kpc, in comparison to 23\%
(14/60) at $d_{\rm em}>1$ kpc, confirming a slightly higher detection rate for the compact sources. However, this is
elevated by the 0.1 -- 1 kpc bin. Furthermore, if the linear extent of the source is correlated with its luminosity, in
conjunction with the fact that the radio and ultra-violet luminosities are correlated \citep{ace+12}, we would expect a
lower detection rate for the large/more luminous sources, due to ionisation of the gas by the active galactic nucleus
\citep{cw12}.  When sources above the critical luminosity ($L_{\rm UV} \sim10^{23}$ \WpHz) are removed, 21-cm detection rates
in compact objects are not significantly higher than for the rest of the population \citep{cw10}.
%\footnote{Above this ultra-violet ($\lambda=912$ \AA) luminosity all of the gas is believed to be ionised by the active galactic nucleus \citep{cw12}.}

%\subsection{Modelling a double lobed radio source}
\subsection{Absorption of a double lobed radio source}

\begin{figure*}
\centering
\hspace*{-3mm}\includegraphics[angle=0,scale=0.8]{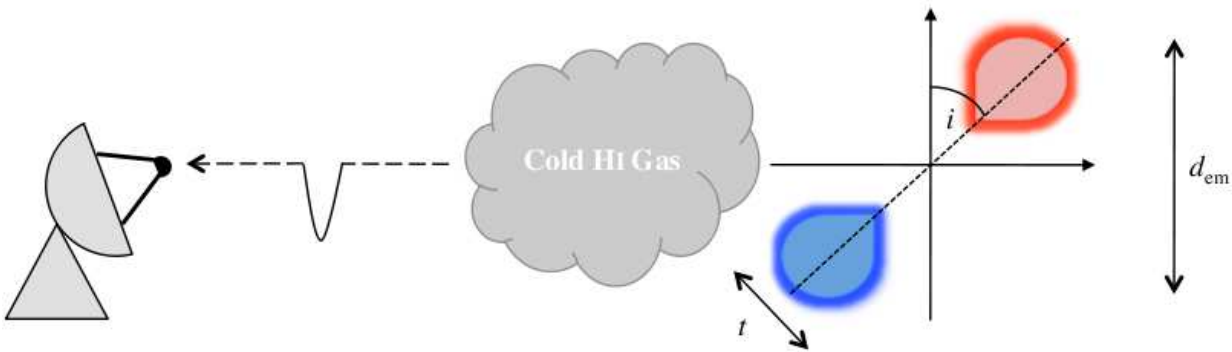}
\caption{Schematic of the model showing projected linear size, $d_{\rm em}$ in relation to the lobe inclination, $i$, and thickness, $t$ \citep{glo12}.}
\label{geom}
\end{figure*}

\begin{figure}
\centering \includegraphics[angle=0,scale=0.47]{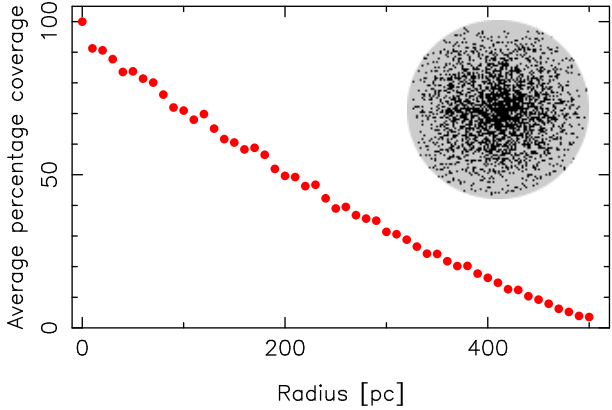}
\caption{A single realisation of the average coverage with radius for a 1~kpc non-uniform absorber. The inset shows the density distribution in the RA--$\delta$ plane.}
\label{profile}
\end{figure}

\begin{figure}
\centering \includegraphics[angle=0,scale=0.70]{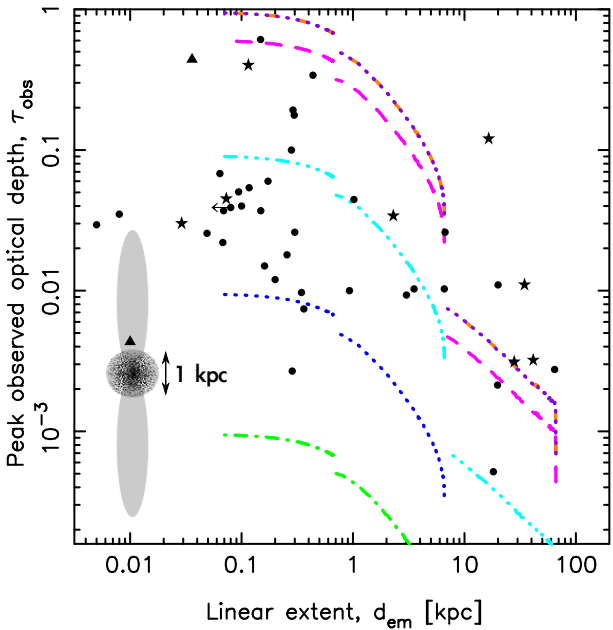} %tau-size_all+1ell.ps}
\caption{The observed optical depth versus the projected size (where available)  for all of the 21-cm detections (see Fig. \ref{all}).
The overlain lines show the variation for a non-uniformly dense absorber of $d_{\rm abs} = 1$ kpc, occulting the elliptical lobe model (FR{\sc i}), with
$t = d/10$ for $d = 0.66$, $6.6$ and $66$ kpc (the largest of the detections).
The maximum value on the abscissa shows the deprojected size
and on the ordinate the same values of $\tau_{\rm obs}$ as shown in Fig.~\ref{pcv2}. %This is 
 The inset illustrates the un-inclined 6.6 kpc model in the projected sky plane.}
\label{tau-size}
\end{figure}

\begin{figure}
\centering \includegraphics[angle=0,scale=0.70]{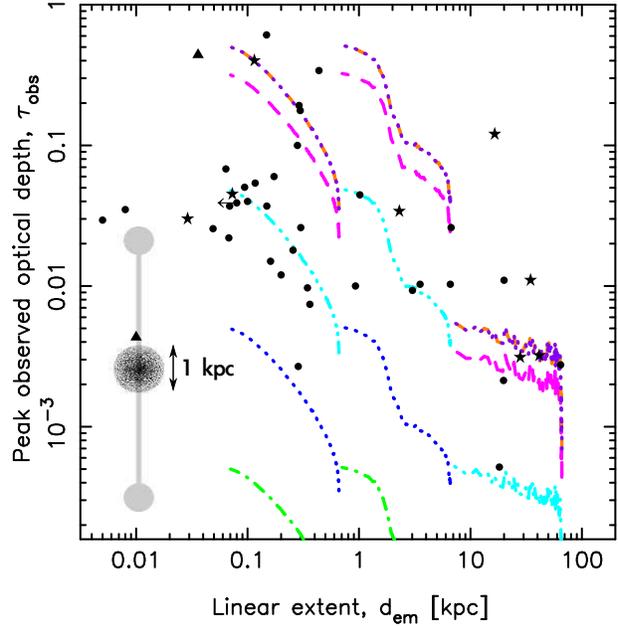}
\caption{As Fig. \ref{tau-size}, but for the lollipop model (FR{\sc ii}), again with a lobe diameter of $t = d/10$, for $d = 0.66$, $6.6$ and $66$ kpc and a ``stick'' width $d/60$.}
\label{tau-stick}
\end{figure}

In order to test whether the wide range in observed source sizes %(Fig. \ref{pcv}) 
could, at least in part, be caused by
projection effects, we model the source size, $d_{\rm em}$, through the inclination, $i$,  of a small range of intrinsic sizes.\footnote{It is the general consensus that,
at least some (CSOs \& CSSs), compact objects are young and therefore intrinsically small (e.g. \citealt{ffd+95}). %,dgs+00,oc98,mff+99}). 
However, since we are
including all of the redshifted associated systems in which 21-cm has been detected, it is of interest to see if the inclination of a single intrinsic size can account for
the observed range in projected sizes.}
% In any case, whether the projected size is intrinsic or caused by orientation has no bearing on the covering  factor along our line-of-sight.  {\color{red}{{\tt is this right?}}}}
That is, $d_{\rm em} = t + d\cos i$, where
$t$ is the diameter of the lobe and $d$ the deprojected space between the lobe centres (Fig. \ref{geom}).
As described by \citet{glo12}, the cross-sectional area of the lobes %, $A_{\rm em}$, 
intercepted by a spherical absorber, %(of cross-sectional area $A_{\rm abs} = {\rm \pi}d{_{\rm abs}}^2/4$)
i.e. the covering factor, was recorded when 
incrementing the inclination. %Since the observed data show considerable scatter (Fig. \ref{all}), most likely due to various intrinsic optical depths, 
Although there may be a preferred absorber size (Sect. \ref{ext}), we do not know the underlying distribution of the intrinsic properties (such as optical 
depth and deprojected size) and so cannot perform a reliable statistical fit to these data. % (e.g. the least-squares fit in Fig. \ref{all}).
Therefore, for various intrinsic optical depths, $\tau$, the covering factor was converted to the observed optical depth via  $\tau_{\rm obs} = f(1-e^{-\tau})$,
with the best fit to the trend of the detections judged by eye.
Different models were tested through the variation of several parameters: % \citep{glo12}:
\begin{itemize}
  \item[-] Varying the relative sizes of the model components: Specifically, the size of the absorber in relation to the lobe size, lobe thickness and
                 spacing between the lobes, as well variation of the relative sizes between these three parameters. Generally, an absorber size close to or larger than the
deprojected spacing between the lobes gave the better fit. % when this was larger than the lobe separation. 
% A small absorber size in relation to the separation does not intercept any emission.

\item[-] Varying the lobe morphologies: Several morphologies were tested, with the best results being obtained by an 
ellipsoid\footnote{Which was projected as ellipse that shortened with inclination until forming a circle at $i=90$\dg.}, modelling the
dominant jet morphology of an FR{\sc i} source \citep{fr74}, and a ``lollipop'',\footnote{In which the ``stick'' shortened with inclination until forming a circle at $i=90$\dg.}
modelling the faint jet plus hot-spot morphology of an FR{\sc ii} source.
% \begin{figure}
% \centering \includegraphics[angle=0,scale=0.85]{lobe_and_stick_model.eps}
% \caption{The model ``lollipop''  (FR{\sc ii} morphology), where the lobe consists of a jet and a localised hot-spot. {\color{red}{{\tt JAMES - make more like
%       Marcin's lollipop (according to the parameters he will hopefully send and show axes indicating view is from front (cf. the side in Fig. \ref{geom})}}} THAT IS, IF WE KEEP IT - might not be
% necessary and we may not have space}
% \label{stick}
% \end{figure}
\item[-] Introducing a non-uniform density distribution: In order to simulate the clumpy nature of the absorber, the
  points forming the density distribution were generated % from a uniform probability distribution, where if a randomly generated 
randomly, where if a value (RA$_n$,$\delta_n$) is at a radius smaller than the next generated value 
(RA$_{n+1}$,$\delta_{n+1}$), then absorption occurs at  (RA$_n$,$\delta_n$). As seen from Fig. \ref{profile},
this yields a clumpy distribution in which the density of absorbing points falls with radius.
% with the spread dictated by the radius of the absorber (see \citealt{glo12}). 
Using the non-uniform absorber has the effect of smoothing the dependence
  of the observed optical depth on inclination, generally providing a better match to the data than a uniformly dense
  absorber.
\end{itemize}
In Figs. \ref{tau-size} and \ref{tau-stick} we show distributions obtained for the FR{\sc i} and FR{\sc ii} approximations, respectively, intercepted by a non-uniformly
dense absorber of total extent 1 kpc (Fig. \ref{profile}), which yielded the $\tau_{\rm obs} - d_{\rm em}$ slopes that best traced the observed distribution \citep{glo12}.
From these, just as for the intrinsic optical depths (Sect. \ref{odcf}), the projection of a single intrinsic source size cannot account for the whole distribution (as found for the compact objects by
\citealt{ffs+90}), although
a range of sizes in conjunction with a range of optical depths does exhibit a similar slope to the $\tau_{\rm obs} - d_{\rm em}$ anti-correlation.  At large
linear extents, there are two non-compact objects which are located away from the model ``fits'', although these did not form part of the original sample
which led to the perceived anti-correlation (cf. Figs. \ref{pcv2} and \ref{pcv1}). 
 
Lastly, \citet{pcv03} model the absorbing gas as both a spherical cloud and a disk, centered on the nucleus (the origin of the axes in Fig. \ref{geom}). In the former case
the model results will be the same as presented above, since, for a given absorber,  the covering factor is solely dependent on the projected area of the lobes intercepted
by the absorber. In the latter case,  where the absorption is occuring in the pc-scale torus or the kpc-scale galactic disk, with a fixed
orientation perpendicular to the jet axis, we may expect the power law drop as seen by \citet{pcv03}, although this would be with respect to the observed optical
depth and not the column density (cf. Figs. \ref{pcv} and \ref{pcv2}). However, from the incidence of 21-cm detections in type-1 and type-2 active galactic nuclei, \citet{cww+08}
conclude that the bulk absorption occurs in the galactic disk, which must be randomly orientated with respect to the torus. This rules out
applying a generic  model of the jet inclination to a disk distribution, although we would generally expect the observed optical depth to decrease with decreasing inclination due to
 a reduction in the covering factor (Sect. \ref{odcf}).

\section{Discussion and summary}
%\newpage

Knowledge of the density of neutral gas in the distant Universe is crucial to our understanding of its contribution to
the cosmic mass density (e.g. \citealt{cur09a} and references therein) and its relation to the star formation history
(e.g. \citealt{hb06}). Therefore, any dependence of the total column density of the neutral gas has profound consequences in the interpretation of
this and, since \citet{pcv03} noted the column density to be anti-correlated with size of the radio source illuminating the \HI\
cloud, this relationship has been confirmed several times,
%\citep{gs06,gs06a,gss+06,omd06,css11}. However, 
although no clear physical explanation has ever been provided. 

Upon dissecting the components which make up the estimate of the column density, we find, as \citeauthor{pcv03},
the correlation to be between the observed optical depth and radio source size.  The former quantity is directly
proportional to the covering factor, in the optically thin regime, which is itself inversely proportional to the source
size.  Therefore, these quantities are not independent, with an observed optical depth--source size
anti-correlation expected purely from the definition of $\tau_{\rm obs}$ alone.  It is the assumption of a
common spin temperature and covering factor for all of the sources, which leads to the $N_{\rm HI} - d_{\rm em}$
anti-correlation.
 
Extending the sample to include the non-compact objects, as found by \citet{dob+98,fpf+00}, we find no distinct difference
between these and the compact sources, with the non-compact objects also following the $\tau_{\rm obs} - d_{\rm em}$
anti-correlation. Also, while the 21-cm detection rate may be higher in compact objects (e.g. \citealt{gss+06}), this
could be dominated by a ``resonance'' at $d_{\rm em}\approx0.1 - 1$ kpc, in addition to the fact that the larger sources
may have higher ultra-violet luminosities, decreasing the likelihood of detecting neutral gas \citep{cw10}. This
resonance, where the 21-cm detection rate is significantly higher than for other values of $d_{\rm em}$, may indicate
that the typical cross section of cold absorbing gas is of the order $100 -1000$ pc. %, which is supported by our model.

Finally, through a model of a double lobed radio source, where the projected linear size, $d_{\rm em}$, depends upon the
inclination, we find that the anti-correlation can be best accounted for models which approximate either of the two
Fanaroff \& Riley classifications and a non-uniformly dense absorber with a diameter in the range of the resonant
$d_{\rm em}$.
 We also confirm the finding of \citet{ffs+90} that the observed range in source extents cannot be accounted for by the inclination of a single deprojected size.

\section*{Acknowledgements}
 
This research was conducted by the Australian Research Council Centre of Excellence for All-sky Astrophysics (CAASTRO), through project number CE110001020.
JRA acknowledges support from an Australian Research Council Super Science Fellowship.
%\bibliographystyle{../mn2e}   
%\bibliography{aa,ref}

\begin{thebibliography}{34}
\expandafter\ifx\csname natexlab\endcsname\relax\def\natexlab#1{#1}\fi

\bibitem[{{Allison} {et~al}\mbox{.}(2012){Allison}, {Curran}, {Emonts},
  {Gereb}, {Mahoney}, {Reeves}, {Sadler}, {Tanna}, \& {Whiting}}]{ace+12}
{Allison} J.~R. {et~al.}, 2012, MNRAS, 423, 2601

\bibitem[{{Braun}(2012)}]{bra12}
{Braun} R., 2012, ApJ, 87, 749

\bibitem[{{Chandola} {et~al}\mbox{.}(2011){Chandola}, {Sirothia}, \&
  {Saikia}}]{css11}
{Chandola} Y., {Sirothia} S.~K., {Saikia} D.~J., 2011, MNRAS, 418, 1787

\bibitem[{Curran(2010)}]{cur09a}
Curran S.~J., 2010, MNRAS, 402, 2657

\bibitem[{Curran(2012)}]{cur12}
Curran S.~J., 2012, ApJ, 748, L18

\bibitem[{Curran {et~al}\mbox{.}(2007)Curran, Tzanavaris, Pihlstr\"{o}m, \&
  Webb}]{ctp+07}
Curran S.~J., Tzanavaris P., Pihlstr\"{o}m Y.~M., Webb J.~K., 2007, MNRAS, 382,
  1331

\bibitem[{Curran \& Whiting(2010)}]{cw10}
Curran S.~J., Whiting M.~T., 2010, ApJ, 712, 303

\bibitem[{Curran \& Whiting(2012)}]{cw12}
Curran S.~J., Whiting M.~T., 2012, ApJ, 759, 117

\bibitem[{Curran {et~al}\mbox{.}(2011{\natexlab{a}})Curran, Whiting, Murphy,
  Webb, Bignell, Polatidis, Wiklind, Francis, \& Langston}]{cwm+10}
Curran S.~J. {et~al.}, 2011{\natexlab{a}}, MNRAS, 413, 1165

\bibitem[{Curran {et~al}\mbox{.}(2013{\natexlab{a}})Curran, Whiting, Sadler, \&
  Bignell}]{cwsb12}
Curran S.~J., Whiting M.~T., Sadler E.~M., Bignell C., 2013{\natexlab{a}},
  MNRAS, 428, 2053

\bibitem[{Curran {et~al}\mbox{.}(2013{\natexlab{b}})Curran, Whiting, Tanna,
  Sadler, \& Athreya}]{cwt+12}
Curran S.~J., Whiting M.~T., Tanna A., Sadler E.~M., Athreya R.,
  2013{\natexlab{b}}, MNRAS, 429, 3402

\bibitem[{Curran {et~al}\mbox{.}(2011{\natexlab{b}})Curran, Whiting, Webb, \&
  Athreya}]{cwwa11}
Curran S.~J., Whiting M.~T., Webb J.~K., Athreya A., 2011{\natexlab{b}}, MNRAS,
  414, L26

\bibitem[{Curran {et~al}\mbox{.}(2008)Curran, Whiting, Wiklind, Webb, Murphy,
  \& Purcell}]{cww+08}
Curran S.~J., Whiting M.~T., Wiklind T., Webb J.~K., Murphy M.~T., Purcell
  C.~R., 2008, MNRAS, 391, 765

\bibitem[{{de Vries} {et~al}\mbox{.}(1998){de Vries}, {O'Dea}, {Perlman},
  {Baum}, {Lehnert}, {Stocke}, {Rector}, \& {Elston}}]{dob+98}
{de Vries} W.~H., {O'Dea} C.~P., {Perlman} E., {Baum} S.~A., {Lehnert} M.~D.,
  {Stocke} J., {Rector} T., {Elston} R., 1998, ApJ, 503, 138

\bibitem[{{Fanaroff} \& {Riley}(1974)}]{fr74}
{Fanaroff} B.~L., {Riley} J.~M., 1974, MNRAS, 167, 31P

\bibitem[{{Fanti}(2000)}]{fan00}
{Fanti} C., 2000, in EVN Symposium 2000, Proceedings of the 5th european VLBI
  Network Symposium, {Conway} J.~E., {Polatidis} A.~G., {Booth} R.~S.,
  {Pihlstr{\"o}m} Y.~M., eds., p.~73

\bibitem[{{Fanti} {et~al}\mbox{.}(1995){Fanti}, {Fanti}, {Dallacasa},
  {Schilizzi}, {Spencer}, \& {Stanghellini}}]{ffd+95}
{Fanti} C., {Fanti} R., {Dallacasa} D., {Schilizzi} R.~T., {Spencer} R.~E.,
  {Stanghellini} C., 1995, A\&A, 302, 317

\bibitem[{{Fanti} {et~al}\mbox{.}(2000){Fanti}, {Pozzi}, {Fanti}, {Baum},
  {O'Dea}, {Bremer}, {Dallacasa}, {Falcke}, {de Graauw}, {Marecki}, {Miley},
  {Rottgering}, {Schilizzi}, {Snellen}, {Spencer}, \& {Stanghellini}}]{fpf+00}
{Fanti} C. {et~al.}, 2000, A\&A, 358, 499

\bibitem[{{Fanti} {et~al}\mbox{.}(1990){Fanti}, {Fanti}, {Schilizzi},
  {Spencer}, {Nan Rendong}, {Parma}, {van Breugel}, \& {Venturi}}]{ffs+90}
{Fanti} R., {Fanti} C., {Schilizzi} R.~T., {Spencer} R.~E., {Nan Rendong},
  {Parma} P., {van Breugel} W.~J.~M., {Venturi} T., 1990, A\&A, 231, 333

\bibitem[{Glowacki(2012)}]{glo12}
Glowacki M., 2012, {The Effects of Geometry on the Apparent Star-Forming Gas
  Abundance in Radio Galaxies}. Tech. rep., University of Sydney

\bibitem[{{Gupta} \& {Saikia}(2006{\natexlab{a}})}]{gs06}
{Gupta} N., {Saikia} D.~J., 2006{\natexlab{a}}, MNRAS, 370, L80

\bibitem[{{Gupta} \& {Saikia}(2006{\natexlab{b}})}]{gs06a}
{Gupta} N., {Saikia} D.~J., 2006{\natexlab{b}}, MNRAS, 370, 738

\bibitem[{{Gupta} {et~al}\mbox{.}(2006){Gupta}, {Salter}, {Saikia}, {Ghosh}, \&
  {Jeyakumar}}]{gss+06}
{Gupta} N., {Salter} C.~J., {Saikia} D.~J., {Ghosh} T., {Jeyakumar} S., 2006,
  MNRAS, 373, 972

\bibitem[{{Gupta} {et~al}\mbox{.}(2012){Gupta}, {Srianand}, {Petitjean},
  {Bergeron}, {Noterdaeme}, \& {Muzahid}}]{gsp+12}
{Gupta} N., {Srianand} R., {Petitjean} P., {Bergeron} J., {Noterdaeme} P.,
  {Muzahid} S., 2012, A\&A, 544, 21

\bibitem[{{Hopkins} \& {Beacom}(2006)}]{hb06}
{Hopkins} A.~M., {Beacom} J.~F., 2006, ApJ, 651, 142

\bibitem[{Kanekar \& Chengalur(2003)}]{kc02}
Kanekar N., Chengalur J.~N., 2003, A\&A, 399, 857

\bibitem[{{Lavalley} {et~al}\mbox{.}(1992){Lavalley}, {Isobe}, \&
  {Feigelson}}]{lif92a}
{Lavalley} M.~P., {Isobe} T., {Feigelson} E.~D., 1992, in BAAS, Vol.~24, pp.
  839--840

\bibitem[{{O'Dea}(1998)}]{ode98}
{O'Dea} C.~P., 1998, PASP, 110, 493

\bibitem[{{O'Dea} {et~al}\mbox{.}(1994){O'Dea}, {Baum}, \& {Gallimore}}]{obg94}
{O'Dea} C.~P., {Baum} S.~A., {Gallimore} J.~F., 1994, ApJ, 436, 669

\bibitem[{{Orienti} {et~al}\mbox{.}(2006){Orienti}, {Morganti}, \&
  {Dallacasa}}]{omd06}
{Orienti} M., {Morganti} R., {Dallacasa} D., 2006, A\&A, 457, 531

\bibitem[{{Pihlstr{\" o}m} {et~al}\mbox{.}(2003){Pihlstr{\" o}m}, {Conway}, \&
  {Vermeulen}}]{pcv03}
{Pihlstr{\" o}m} Y.~M., {Conway} J.~E., {Vermeulen} R.~C., 2003, A\&A, 404, 871

\bibitem[{{Tzioumis} {et~al}\mbox{.}(2002){Tzioumis}, {King}, {Morganti},
  {Dallacasa}, {Tadhunter}, {Fanti}, {Reynolds}, {Jauncey}, {Preston},
  {McCulloch}, {Tingay}, {Edwards}, {Costa}, {Jones}, {Lovell}, {Clay},
  {Meier}, {Murphy}, {Gough}, {Ferris}, {White}, \& {Jones}}]{tkm+02}
{Tzioumis} A. {et~al.}, 2002, A\&A, 392, 841

\bibitem[{{van Breugel} {et~al}\mbox{.}(1984){van Breugel}, {Miley}, \&
  {Heckman}}]{vmh89}
{van Breugel} W., {Miley} G., {Heckman} T., 1984, AJ, 89, 5

\bibitem[{{Wolfe} \& {Burbidge}(1975)}]{wb75}
{Wolfe} A.~M., {Burbidge} G.~R., 1975, ApJ, 200, 548

\end{thebibliography}

\label{lastpage}

\end{document}